# Fpack and Funpack User's Guide
## FITS Image Compression Utilities

William Pence, NASA Goddard Space Flight Center, Greenbelt, MD 20771
Rob Seaman, National Optical Astronomy Observatory, Tucson, AZ 85719
Rick White, Space Telescope Science Institute, Baltimore, MD 21218

February 2011

## 1. Introduction

Fpack is a utility program for optimally compressing images in the FITS (Flexible Image Transport System) data format (see http://fits.gsfc.nasa.gov). The associated funpack program restores the compressed image file back to its original state (if a lossless compression algorithm is used). (An experimental method for compressing FITS binary tables is also available; see section 7). These programs may be run from the host operating system command line and are analogous to the gzip and gunzip utility programs except that they are optimized for FITS format images and offer a wider choice of compression options.

Fpack stores the compressed image using the FITS tiled image compression convention (see http://fits.gsfc.nasa.gov/fits_registry.html). Under this convention the image is first divided into a user-configurable grid of rectangular tiles, and then each tile is individually compressed and stored in a variable-length array column in a FITS binary table. By default, fpack usually adopts a row-by-row tiling pattern.

The tiled image compression convention can in principle support any number of different compression algorithms. The fpack and funpack utilities call on routines in the CFITSIO library (http://heasarc.gsfc.nasa.gov/fitsio) to perform the actual compression and uncompression of the FITS images. Currently, the GZIP, Rice, H-compress, and the PLIO IRAF pixel list compression algorithms are supported.

The fpack and funpack utilities were originally designed and written by Rob Seaman (NOAO). William Pence (NASA) added further enhancements to the utilities and to the image compression algorithms in the underlying CFITSIO library. Rick White (STScI) developed the code for the Rice and Hcompress algorithms.

## 2. Benefits of using fpack

Using fpack to compress FITS images offers a number of advantages over the other commonly used technique of externally compressing the whole FITS file with gzip:

1. fpack generally offers higher compression ratios and faster compression speed than GZIP.
2. The FITS image header keywords remain uncompressed and can be read or written without any additional overhead.
3. Each HDU of a multi-extension FITS file is compressed separately, thus it is not necessary to uncompress the entire file to read a single image in a multi-extension file.

4. The capability of dividing the image up into tiles before compression enables faster access to small subsections of the image because only those tiles contained in the subsection need be uncompressed.
5. The compressed image is itself a valid FITS file and can be manipulated by other general FITS utility software.
6. Fpack supports lossy compression techniques that can achieve significantly higher compression factors than the lossless compression algorithms in situations where it is not necessary to exactly preserve every bit of the original image pixel values. This is especially relevant when compressing 32-bit floating point FITS images for which there is often little justification for preserving the full numerical precision (6 - 7 decimal places) of each pixel value.
7. Fpack and Funpack automatically update the CHECKSUM keywords in the compressed and uncompressed files to help verify the integrity of the FITS files.
8. Software applications that are built on top of FITS access libraries such as CFITSIO, that internally support the tiled image compression technique, are able to directly read and write the FITS images in their compressed form, thus reducing the amount of disk storage space needed by users.

Data providers can minimize the data storage and network bandwidth resources needed to archive and distribute FITS images by compressing them first with fpack. Users can then uncompress the FITS with funpack to convert them back into the standard FITS image format before doing any further analysis. The benefits of using fpack are magnified, however, when the analysis software is capable of directly reading and writing the files in the compressed form. This reduces both the required amount of local disk space and the time needed to copy the files from one location to another.

Any software application that uses the CFITSIO library (http://heasarc.gsfc.nasa.gov/fitsio) to read and write FITS images will inherit the ability to read or write tile-compressed images. The image compression or uncompression is performed internally by the CFITSIO library routines, so in general, the application program itself does not need to know anything about the tiled compressed image format. The main exception to this is that when writing compressed images, the application program may need to call an additional routine to define which compression algorithm to use, along with the values of other optional compression parameters. The fpack and funpack utilities are themselves examples of applications that use CFITSIO to perform the compression and uncompression operations on the images.

Besides CFITSIO, the ds9 image display program and the IRAF data analysis system currently provide some support for the tile-compressed FITS image format. It is anticipated that other analysis systems will also add support for this tiled image compression format as it become more widely used. In the meantime, funpack may be used to uncompress the images back into standard FITS images for compatible with other astronomical image analysis software that does not yet directly support the compressed format.

## 3. Compression versus noise

When losslessly compressing images, the amount of compression that can be achieved depends almost completely on one simple factor: the amount of the noise that is present in the pixel values. The noise, by definition, cannot be compressed, so the compression ratio of an image will be inversely proportional to the total number of noise bits in the image. As is discussed in greater detail in our Paper I (Pence, Seaman, and White, 2009, PASP 121,414; preprint: http://arxiv.org/abs/0903.21401) the amount of noise in a image can be calculated from the measured standard deviation (sigma) of the pixels in the "background" areas of the image (e.g., excluding bright stars of other objects in the image) which typically closely approximates a Gaussian distribution. The average number of noise bits per pixel is then given by

$$\text{Nbits} = \log_2(\text{sigma}) + 1.792$$

Since these noise bits cannot be compressed, the maximum possible compression ratio, in the ideal case where all the remaining bits are compressed to zero, is simply given by the ratio BITPIX / Nbits (where BITPIX is the number of bits in each pixel value). No actual compression algorithm can achieve this theoretical limit, so in practice the compression ratio is given by BITPIX / (Nbits + K) where K is a measure of the efficiency of the particular compression algorithm. For the Rice algorithm, K has a value of about 1.2, and for Hcompress it is about 0.9. The k value for GZIP is much larger (and variable), typically about 4 or 5.

The statistical noise in each image pixel value scales with the square root of the number of detected photons. (There may be other sources of noise as well). Thus, the amount of noise in the image naturally increases with the mean count rate, or exposure time. The practical implication of this fact is that the different types of exposures taken during a typical astronomical observing session (e.g., in order of increasing mean count rate: bias frames, short calibration exposures, deep sky exposures, and flat field images) will have distinctly different amounts of noise and hence will compress by differing amounts. Thus, bias frames, with the lowest pixel count values, will compress better than flat field images or long exposures of the night sky that have much larger pixel values.

## 4. Lossless compression of integer FITS images.

In Paper I, we used a large set of direct imaging CCD exposures from NOAO taken of star fields in the night sky, plus the associated calibration exposures, to compare the compression speeds and file compression ratios for the 3 different general purpose compression algorithms that are currently supported by fpack, namely, Rice, GZIP, and Hcompress. We also compared these to the widely-used method of compressing the entire FITS file with the host-level GZIP file compression program.

The mean file compression ratios and the relative compression and uncompression elapsed CPU times for these 4 different compression methods are shown in Table 1. These values are the mean for all 1632 16-bit integer images in the sample data set; the CPU times in each case are relative to those when using the Rice algorithm.

Table 1. Compression Statistics for 16-bit Integer Images

|  | Rice | Hcompress | GZIP | Host GZIP |
|---|---|---|---|---|
| **Compression Ratio** | 2.11 | 2.18 | 1.53 | 1.6 |
| **Relative compression CPU time** | 1.0 | 2.8 | 5.6 | 2.6 |
| **Relative uncompression CPU time** | 1.0 | 3.1 | 1.9 | 0.9 |

As shown in the first row, the Rice and Hcompress methods achieve significantly larger compression of these astronomical images than GZIP. The GZIP compressed files are on average about 1.4 times larger than the Rice or Hcompressed files. This depends slightly on the amount of noise in the image: the ratio is about 1.3 for the images with the most amount of noise and about 1.5 for the least noisy images. Hcompress produces slightly better compression than Rice (about 3% smaller), but for most applications this small gain is not worth the much greater CPU times required to compress and uncompress the images with Hcompress.

The second and third rows of the table show that the Rice compression algorithm is generally considerably faster than Hcompress or GZIP. Note that the factor of approximately 2 timing difference between using the host-level GZIP program to compress an image and using the tiled-image implementation of the same GZIP algorithm within fpack/CFITSIO is due to the fact that the host-level GZIP program can read and write the files more efficiently as simple continuous streams, whereas the fpack implementation requires that the input and output files be copied to and from intermediate storage buffers in memory. As a benchmark point of reference, a Linux machine with a 2.4 GHz AMD Opteron 250 dual core processor can compress or uncompress a 50 MB 16-bit integer image in 1 second of CPU time when using the Rice compression algorithm.

Similar trends are seen when compressing 32-bit integer images, only the compression factors that are achieved are typically twice that of a 16-bit image, given the same noise level. See Paper I for more details.

## 5. Compression of floating-point FITS images

It is generally not practical (nor necessary) to losslessly compress floating-point format FITS images (which have BITPIX = -32 or -64). This is because many of the compression algorithms only support integer data, not floating-point, and even if they do (e.g., GZIP), a large fraction of the bits in the mantissa of the image pixel values are often filled with uncompressible noise, which severely reduces the file compression ratios. A 32-bit floating-point value can represent 6 – 7 decimal places of precision, which usually far exceeds the accuracy of individual image pixel values. For this reason, fpack usually quantizes the pixel values into 32-bit integers using a linear scaling function:

$$\text{integer\_value} = (\text{floating\_point\_value} - \text{ZERO\_POINT}) / \text{SCALE\_FACTOR}$$

This array of scaled integers is then compressed using one of the supported compression algorithms (usually Rice). When the image is subsequently uncompressed, the integer values are inverse scaled to closely, but not exactly, reproduce the original floating point pixel values. Separate scale and zero point values are computed for each tile of the image. These and other issues related to compressing floating-point images are discussed in greater detail in a our Paper II (Pence, White, & Seaman 2010, PASP in press; preprint: http://arxiv.org/abs/1007.1179).

The value of SCALE_FACTOR in the scaling function controls how closely the inverse scaled values approximate the original floating point values: decreasing SCALE_FACTOR reduces the spacing between the quantized levels in the inverse-scaled values and thus more closely reproduces the original pixel values. However, this also magnifies the dynamic range and the noise level in the integer array that is to be compressed, which adversely affects the amount of compression that is achieved. Thus, there is a direct trade-off between providing more precision or achieving greater compression.

One refinement to the quantization procedure described above is to add a small amount of random noise to the floating point value before scaling it to an integer; that same random value is then subtracted when converting back to the floating point value. This "subtractive dithering" technique helps preserve low amplitude signals in the quantized image through an effect known as "stochastic resonance". This is especially important for preserving the mean value of the background sky level when measuring the flux of faint sources in the image.

It is not easy to directly determine an appropriate SCALE_FACTOR value to use with a given image, therefore fpack provides instead a quantization parameter called "q" for specifying how closely the inverse-scaled integer pixel values must approximate the original floating point pixel values, relative to the measured noise in background areas in the image. The image pixel values will be quantized so that the spacing between the adjacent discrete levels is equal to the measured sigma of the R.M.S. noise in the background regions of the image divided by q. The maximum deviation between the pixel values in the compressed image and in the original image will be half this value. Formally, the number of noise bits that are preserved in each pixel value is given by $\log_2(q) + 1.792$. For example, if q = 4, then the quantized levels are spaced at intervals of sigma/4 and about 3.8 bits of noise are preserved in each pixel value. Increasing the value of q will produce compressed images that more closely approximate the pixel values in the original floating point image, but will also increase the size of the compressed image file, as shown in Table 2. The third column in this table shows how much the background noise will increase in the quantized images, which is given by $\sqrt{1 + 1/(12q^2)}$. The numerical experiments described in Paper II demonstrate that the statistical noise on the measured magnitudes and positions of faint stars in the quantized images can be expected to increase by the same percentage amount.

Table 2. Compression of Floating Point Images

| q | Compression Ratio | Noise increase % |
|---|---|---|
| 1 | 9.5 | 4.1 |
| 2 | 8 | 1 |
| 4 | 6.5 | 0.26 |
| 8 | 5.3 | 0.07 |

fpack users are urged to perform quantitative tests on there own data sets using different values of q to determine the appropriate value for their particular application.

In rare situations, or for test purposes, it may be desirable to losslessly compress floating point FITS images. This can be accomplished by specifying the GZIP compression algorithm with -g, -g1, or -g2 (because the other algorithms can only compress integers) and a q value of 0 (e. g., "fpack -g -q 0"). This will exactly preserve every bit in the floating-point image at the expense of much lower compression ratios than given by the quantization technique.

## 6. Lossy compression of integer FITS images

Some integer FITS images contain too much noise to be losslessly compressed very effectively. In our Paper I, for example, Figure 3 shows a case of deep CCD exposures which contain 7 – 8 bits of noise per pixel that can only be losslessly compressed by less than a factor of 2. Just as with typical floating-point images, removing some of the noise from these integer images can increase the image compression ratio without significantly degrading the scientific accuracy of measurements in the image.

Fpack has an option (specified with the -i2f command line parameter) to internally convert integer FITS images into floating-point format, and then apply the same quantization procedure that it uses for actual floating-point FITS images. The compression ratio will depend primarily on the specified q quantization parameter, as shown in Table 2, except that the compression ratio will only be ½ of the amount shown in the table, since the original 16-bit integer pixels are ½ the size of the 32-bit floating point-pixels. Note that if these compressed images are subsequently uncompressed, they will have a floating-point, not integer, data type.

If an integer image contains too little noise, then it does not make sense to use this lossy compression option because the compressed file will actually be larger than if the standard lossless compression method was used. To prevent this, there are 2 other fpack parameters that specify the minimum threshold for the amount of noise in an image for this lossy compression method to be applied. If these thresholds are not met, then the -i2f option is ignored and the integer image is losslessly compressed. For further information, see the description of the -n3ratio and -n3min fpack parameters in the following section.

## 7. Compression of FITS binary tables (BETA feature)

An experimental capability to losslessly compress FITS binary tables has been added to fpack beginning with version 3.26. This uses a prototype FITS convention (http://fits.gsfc.nasa.gov/tiletable.pdf) for compressing each column of the table and storing the compressed bytes in a variable-length array column in the compressed table. This new feature has been added to fpack to facilitate further experiments into the effectiveness of this potential compression method. It is not intended for actual use by projects for publicly distributed data. There is no guarantee at this time that future versions of fpack and funpack will continue to support this table compression format.

This table compression method is invoked by adding the -BETAtable option on the command line. Funpack will automatically uncompress these tables (no command line switch is required). By default, fpack may try several different compression methods for each column, depending on the data type of the column, and then use the one that produces the greatest amount of compression. Alternatively, the -fast option may be specified, to force fpack to simply use the compression method that is likely (based on previous experiments) to produce the best compression for that type of column.

## 8. fpack command-line parameters

The fpack program is invoked on the command line like other host-level utility programs:

```
fpack [OPTION]... [FILE]...
```

The specified options must appear before the list of files to be compressed. The file names may contain the usual wildcard characters that will be expanded by the Unix (or Windows) shell. The input FITS file can be read from the `stdin` file stream by specifying a hyphen as the file name.

The compression algorithm is selected with one of the following options:

- **-r**          Rice [default], or
- **-h**          Hcompress, or
- **-g** or **-g1**   GZIP (per-tile), or
- **-g2**         GZIP (per-tile) (shuffle the bytes in the pixel array first)
- **-p**          IRAF pixel list compression algorithm. This can only be applied to images whose pixel values all lie in the range 0 to $2^{24}$ (16777216).
- **-d**          no compression (debugging mode)

Tiling pattern specification:

When using the Rice, GZIP, or PLIO compression algorithms, the default tiles each contains 1 row of the image (i.e., the tiles are one dimension and contain NAXIS1 pixels). The Hcompress algorithm requires that the tiles be 2-dimensional, therefore the default is to use 16 rows of the image per tile. If this would cause the last partially full tile of the image to only contain a small number of rows, then the tile size is adjusted so that the last tile is more equal in size to the other tiles. The default tile sizes can be overridden with one of the following fpack options:

- **-w**          compress the whole image as a single large tile
- **-t <axes>**   comma separated list of tile dimensions (e. g., -t 200,200 will produce tiles that are 200 x 200 pixels in size)

Compression parameters for floating point images:

**-q <level>** (default value = 4)   Quantization level when compressing floating point images. See the previous discussion on compressing floating point images for more information on this parameter. It is important to realize that fpack does not exactly preserving the original pixel values when compressing floating point images. Users should carefully evaluate the compressed images (e.g., by uncompressing them with funpack) to make sure that any essential information in the original image has not been lost.

Positive q values are interpreted as relative to the R.M.S. noise in the image. A larger q value will more closely preserve the original pixel values, and will result in less compression (see Table 2). In some instances it may be desirable to specify the exact q value (not relative to the measured noise), so that all the tiles in the image, and all the images in a dataset, are compressed using the identical value, regardless of slight variations in the measured noise level. This can be done by specifying the negative of the desired value. In this case, smaller absolute values of q (e.g., -.001 instead of -.01), will better preserve the original pixel values, at the expense of smaller (worse) compression ratios. The -T option (described below) can be used to calculate the noise level in the image.

By default, 'subtractive dithering' is applied to the quantized and scaled values to better preserve faint features in the image (see Section 6). In order to avoid applying the exact same dithering pattern to every image (which is not desirable if one is stacking many compressed images or taking the difference between images) one of 10000 different starting points in the dithering sequence is randomly chosen based on a 'seed' value. There are several ways to compute this seed value:

(1) By default, the seed is computed from the system clock at run time. This means that the quantized pixel values will be slightly different each time an image is compressed.
(2) If one specifies -qt instead of -q, then the seed will be computed from the checksum value of the first tile of image pixels that is scaled and compressed. With this option the seed value will be constant and reproducible for a given image, however, if 2 images have the same pixel values in the first tile (e.g., if there is a blank border around the image and the pixels are all equal to zero) then both images will have the same seed value and will use the same dithering pattern.
(3) Lastly, one can specify which of the 10000 dithering patterns to use by appending a number between 1 and 10000, as in "-q3000 4".
(4) On can suppress the subtractive dithering altogether by specifying -q0. This option is not recommended for general use.

**-n <noise>**   This rarely used parameter (the -i2f parameter offers a better compression method in many cases) rescales the pixel values in a previously scaled image to improve the compression ratio by reducing the R.M.S. noise in the image. This option is intended for use with images that use scaled integers to represent floating point pixel values, and in which the scaling was chosen so that the range of the scaled integer values covers the entire allowed range for that integer data type (e.g., -32768 to +32767 for 16-bit integers and -2147483648 to +2147483647 for 32-bit integers). The measured R.M.S. noise in these integer images is typically so huge that they cannot be effectively compressed. This -n option rescales the pixel values so that the R.M.S. noise will be equal to the specified value. Appropriate values of n will likely be in the range from 2 (for low precision and the high compression) to 16 (for the high precision and lower compression). Users should read the section on compressing floating point images, above, for guidelines on choosing an appropriate value for n that does not lose significant information in the image.

Parameters for lossy compression of integer images:

**-i2f**   This parameter (which stands for "integer to float") forces fpack to internally convert images with integer pixel values into floating-point pixels, which are then compressed using the quantization method that is normally used for actual floating-point FITS images. This lossy compression method may achieve significantly higher compression, without significant loss of information, especially for long exposure images that have relatively large number of counts per pixel (and hence a large amount of Poissonian noise). This parameter will be ignored if the amount of noise in the image is less than the absolute and relative noise thresholds set by the -n3min and -n3ratio parameters.

**-n3min <noise>**   (default value = 6.)  This parameter specifies the minimum value of the RMS

background noise in the image in order for the -i2f parameter to take effect. If the image has less noise, then the -i2f parameter is ignored and the integer image is losslessly compressed.

**-n3ratio <ratio>** (default value = 1.2) This parameter specifies the minimum ratio of the RMS background noise in the image divided by the q parameter value. If the image has less noise, then the -i2f parameter is ignored and the integer image is losslessly compressed.

**-s <scale>** Scale factor for lossy compression when using Hcompress. The default value is 0 which implies lossless compression. Positive scale values are interpreted as relative to the R.M.S. noise in the image. Scale values of 1.0, 4.0, and 10.0 will typically produce compression factors of about 4, 10, and 25, respectively, when applied to 16-bit integer images. In some instances it may be desirable to specify the exact scale value (not relative to the measured noise), so that all the tiles in the image, and all the images in a dataset, are compressed with the identical scale value, regardless of slight variations in the measured noise level. This is done by specifying the negative of the desired value (e.g. -30., which would be equivalent to specifying a scale value of 2.0 in an image that has RMS noise = 15.).

It is important to realize that this option achieves the high compression ratios at the expense of not exactly preserving the original pixel values in the image. Users should carefully evaluate the compressed images (e.g., by uncompressing them with funpack) to make sure that any essential information in the image has not been lost.

The compressed output file name is usually constructed by appending ".fz" to the input file name, and the input file is not deleted, but this behavior may be modified with the following parameters:

**-F** force the input file to be overwritten by the compressed file with the same name. This is only allowed when a lossless compression algorithm is used.
**-D** delete the input file after creating the compressed output file.

**-Y** suppress the prompts to confirm the -F or -D options
**-S** output the compressed FITS file to the STDOUT stream (to be piped to another task)

Parameters for compressing FITS binary table extensions

**-BETAtable** This turns on the table compression feature in fpack (images are also compressed)
**-fast** When used with -BETAtable, this may improve the compression speed of fpack, at the expense of some loss in compression factor.
**-T** When used with -BETAtable, produces a report showing the compression ratio for each column and for the table as a whole. The input file remains unchanged and is not compressed.

Other miscellaneous parameters:

**-v** verbose mode; list each file as it is processed
**-L** list all the extensions in all the input, files. No compression is performed.
**-C** don't update FITS checksum keywords

- **-H**   display a summary help file that describes the available fpack options
- **-V**   display the program version number
- **-T**   produce a report that compares the compression ratio and the compression and uncompression times for each of the main compression algorithms.  The input file remains unchanged and is not compressed.  The format of this report is shown below.
- **-R \<filename\>**   write the comparison test report (produced by the -T option) to a file in a format that is suitable for further analysis.

```
 File: ct655046_13.fits
 Ext BITPIX Dimens. Nulls    Min    Max     Mean   Sigma   Noise3 Nbits   MaxR
   0   16  (1112,4096)    0 -31503  25967 -26679.3 2.5e+03   56.8   7.6   2.10

  Type    Ratio      Size (MB)      Pk (Sec) UnPk Exact ElpN  CPUN   Elp1  CPU1
  Native                                                0.024 0.016 0.013 0.010
  RICE    1.83     9.11 ->    4.98    0.57   0.55 Yes 0.053 0.047 0.045 0.040
  HCOMP   1.85     9.11 ->    4.92    1.91   1.56 Yes 0.175 0.159 0.179 0.162
  GZIP    1.35     9.11 ->    6.73    3.07   1.09 Yes 0.114 0.106 0.108 0.101
  NONE    0.99     9.11 ->    9.18    0.35   0.31 Yes 0.022 0.021 0.015 0.013
```

The first line of the report gives the name of the FITS file; the 3$^{rd}$ line gives the following parameters:
- Ext – extension number within the file (zero based)
- BITPIX – FITS datatype of the image (8, 16, 32, -32 or -64)
- Dimens – image dimensions
- Nulls – number of undefined or null pixels in the image
- Min, Max – the minimum and maximum values in the image
- Mean – mean value of all the non-null pixels
- Sigma – standard deviation of all the non-null pixels
- Noise3 – a measure of the noise in the background regions of the image
- Nbits – number of noise bits per pixel = $\log_2(\text{noise3}) + 1.792$
- MaxR – theoretical maximum possible compression ratio = BITPIX / Nbits

This is followed by a table with the following columns:
- Type – name of compression method, if any
- Ratio - file compression ratio
- Size – uncompressed and compressed sizes of the files, in MB
- Pk – the CPU time in seconds to compress the image with fpack
- UnPk – the CPU time in seconds to uncompress the image with funpack
- Exact – is the compression lossless (i.e., does it exactly preserve the pixel values)?

The following 4 parameters give the measured image read rates, in units of seconds/MB
- ElpN – elapsed time to read the entire image with a single subroutine call
- CPUN – CPU time to read the entire image with a single subroutine call
- Elp1 – elapsed time to read the whole image, one row at a time
- CPU1 – CPU time to read the whole image, one row at a time

The rows in this table correspond to the following cases:

    Native – this just gives the read speed of the input uncompressed image
    Rice – when using the Rice compression algorithm
    Hcomp – when using the Hcompress algorithm
    GZIP – when using the gzip algorithm (within the FITS tiled image compression format)
    None – the image is simply tiled and packed into the FITS tiled image format, without performing any compression on the tiles.

## 9. funpack command-line parameters

funpack shares many of the same parameters as fpack as shown below:

Output file naming parameters:

- **-F**    force the input file to be overwritten by the uncompressed file with the same name. This is only allowed when a lossless compression algorithm is used.
- **-D**    delete the input file after creating the compressed output file.
- **-P <pre>**    prepend the <pre> string to the input file name to generate the name of the uncompressed output file.
- **-O <name>**    used to specify the full name of the uncompressed output file.
- **-S**    output the uncompressed FITS file to the STDOUT stream (to be piped to another task)
- **-Z**    recompress the output file with the host gzip program

Other miscellaneous parameters:

- **-v**    verbose mode; list each file as it is processed
- **-L**    list all the extensions in all the input, files. No uncompression is performed.
- **-C**    don't update FITS checksum keywords
- **-H**    display a summary help file that describes the available funpack options
- **-V**    display the program version number

## 9. Building fpack and funpack

The latest versions of the fpack and funpack C source code and binary executables are available from http://heasarc.gsfc.nasa.gov/fitsio/fpack. The source code is also included in the CFITSIO source file distributions (but not necessarily the latest version) available at http://heasarc.gsfc.nasa.gov/fitsio.

To build the software on unix systems, first download and build the CFITSIO library. If necessary, untar the latest version of the fpack and funpack source code into the CFITSIO directory, overwriting the older version. Then enter the commands

    make fpack
    make funpack

in that directory. This will create the fpack and funpack executable files which may be copied to any other suitable directory (e.g. the local /bin directory).

On Windows PCs, one can manually build the fpack and funpack programs using the Visual C++ compiler with the following command lines (after first building the CFITSIO library following the instructions in the README.win32 file that is distributed with CFITSIO):

```
> cl /MD fpack.c fpackutil.c cfitsio.lib /link setargv.obj
> cl /MD funpack.c fpackutil.c cfitsio.lib /link setargv.obj
```

Note: the "/link setargv.obj" argument is optional. It enables support for the "*" and "?" wildcard characters when specifying the names of the input files to fpack and funpack.